\documentclass[conference]{IEEEtran}
\usepackage{cite}
\usepackage[dvips]{graphicx}
\graphicspath{{../eps/}}
\DeclareGraphicsExtensions{.eps}
\usepackage[cmex10]{amsmath}
\usepackage{listings}
\usepackage{color}
\usepackage{caption3}
\DeclareCaptionFormat{listings} {\parbox{\textwidth}{\vspace{1cm}#1#2#3}}
\usepackage{algorithmic}
\usepackage{algorithm}
\usepackage{array}
\usepackage[caption=false,font=footnotesize]{subfig}
\begin{document}

\title{Dynamic Loop Parallelisation}

\author{\IEEEauthorblockN{Adrian Jackson\IEEEauthorrefmark{1} and
Orestis Agathokleous\IEEEauthorrefmark{1}}
\IEEEauthorblockA{\IEEEauthorrefmark{1}EPCC, 
The University of Edinburgh,\\ Kings Buildings,\\ Mayfield Road, Edinburgh,\\ EH9 3JZ, UK}}
\maketitle

\begin{abstract}
Regions of nested loops are a common feature of High Performance Computing (HPC)
codes. In shared memory programming models, such as OpenMP, these structure are
the most common source of parallelism. Parallelising these structures requires the programmers
to make a static decision on how parallelism should be applied. However,
depending on the parameters of the problem and the nature of the code, static decisions on 
which loop to parallelise may not be optimal, especially as they do not enable the exploitation of 
any runtime characteristics of the execution. Changes to the iterations
of the loop which is chosen to be parallelised might limit the amount of processors
that can be utilised.

We have developed a system that allows a code to make a dynamic choice, at runtime, of 
what parallelism is applied to nested loops.  The system works using a source to
source compiler, which we have created, to perform transformations to user's code automatically,
through a directive based approach (similar to OpenMP).  This approach requires the programmer to 
specify how the loops of the region can be parallelised and our runtime library is then responsible 
for making the decisions dynamically during the execution of the code.

Our method for providing dynamic decisions on which loop to parallelise significantly outperforms 
the standard methods for achieving this through OpenMP (using if clauses) and further optimisations 
were possible with our system when addressing simulations where the number of iterations of the loops 
change during the runtime of the program or loops are not perfectly nested.

\end{abstract}

\IEEEpeerreviewmaketitle

\section{Introduction}
High Performance Computing (HPC) codes, and in particular scientific codes, require parallel 
execution in order to achieve a large amount of performance increase. Depending on the 
underlying parallel platform which is used, programmers use different programming models in order
to achieve parallel execution. In distributed memory systems, the message passing programming
model is the most commonly used approach for applying parallelism in the codes. In shared memory 
systems however, an attractive choice for parallel programming is through OpenMP[16].

The parallelisation of codes with OpenMP is often achieved with loop parallelisation.
As long as the iterations of a loop are independent, they can be distributed to the available
processors of the system in order to execute them in parallel. A programmer is
required to specify a loop that can be parallelised by placing compiler directives before
the loop, resolving any dependency issues between the iterations beforehand.
HPC codes often consist of regions with nested loops of multiple levels. In order to
parallelise these regions, a choice must be made on how parallelism should be applied
on the loops. Even though OpenMP supports a variety of strategies for parallelising
nested loops, only a single one can be used to parallelise the code. 

A static choice however,
cannot exploit any runtime characteristics during the execution of the program.
Changes in the input parameters of the executable which affect the iterations of the 
loops may render the parallelisation decision suboptimal. In addition to this, the iterations of a
loop can change at runtime due to the nature of the code. A common feature of HPC
codes is to organise the data into hierarchies, for example blocks of multi-dimensional
arrays. Depending on the problem, the blocks can have different shapes and sizes.
These parameters affect the loops that are responsible for accessing this data. In some
situations, a static decision has the potential to impose a limitation on the amount of
processors that can be used for the parallel execution of the loops. With the current
trend of chip manufactures to increase the number of cores in the processors in each
generation leading to larger and larger shared memory system being readily available 
to computational scientists on the desktop and beyond, a more dynamic approach must be 
considered for taking such decisions.

This report outlines our investigations into various strategies that can be applied at
runtime in order to make a dynamic decision on how to parallelise a region with nested
loops. Our approach is to try to automatically perform modifications to users code before
compilation in order to enable the code to make these decisions dynamically at runtime.
Specifically, we investigated the possibility of having multiple versions of a loop
within a region of nested loops in order to make a dynamic choice on whether a loop
should be execute sequentially or in parallel.

\section{OpenMP}
OpenMP\cite{openmp} is, arguably, the dominant parallel programming model currently used 
for writing parallel programs for used on shared memory parallel systems. Now at version 3.1, and supported by C and 
FORTRAN, OpenMP operates using compiler directives. The programmer annotates their code specifying 
how it should be parallelised. The compiler then transforms the original code into a
parallel version when the code is compiled. By providing this higher level of abstraction, OpenMP codes tend to
be easier to develop, debug and maintain. Moreover, with OpenMP it is very easy to
develop the parallel version of a serial code without any major modifications.

Whilst there are a number of different mechanisms that OpenMP provides for adding parallel 
functionality to programs, the one that is generally used most often is loop parallelisation.  
This involves taking independent iterations of loops and distributing them to a group of
threads that perform these sets of independent operations in parallel. Since each of the 
threads can access shared data, it is generally straightforward to parallelise
any loop with no structural changes to the program.

\section{Nested Loops}
HPC codes, and particularly scientific codes, deal with numerical computations based
on mathematical formulas. These formulas are often expressed in the form of nested
loops, where a set of computations is applied to a large amount of data (generally stored
in arrays) and parallelisation can be applied to each loop individually. The arrays often
consist of multiple dimensions and the access on the data is achieved with the presence
of nested loops. Furthermore it is not uncommon that the arrangement of the data is
done in multiple hierarchies, most commonly in blocks with multi-dimensional arrays,
where additional loops are require in order to traverse all the data. When such code is
presented, a choice must be made on which loop level to parallelise (where the parallelisation
should occur)\cite{Duran04runtimeadjustment}.  A summary of the available strategies is presented 
in Table \ref{tab:nestedloops}

\begin{table}
\renewcommand{\arraystretch}{1.3}
\caption{Strategies for parallelising nested loop regions}
\label{tab:nestedloops}
\centering
\begin{tabular}{||p{0.1\textwidth}|p{0.3\textwidth}||}
\hline
{\bf Name} & {\bf Description}\\
\hline
Outermost & Loop Parallelisation of the outermost loop\\
Inner Loop & Parallelisation of one of the inner loops\\
Nested & Parallelisation of multiple loops with nested parallel regions\\
Loop Collapsing & Collapsing the loops into a single big loop\\
Loop Selection & Runtime loop selection using if clauses \\
\hline
\end{tabular}
\end{table}

\subsection{Outermost loop}

The most commonly used approach is to parallelise the outermost loop of a nested
loop region, as shown in Listing \ref{alg:outerloop}. Using this strategy, the iterations of the loop
are distributed to the members of the thread team. The threads operate in parallel by
executing the portion of iterations they are assigned to them individually. The nested
loops of the parallel region are executed in a sequential manner.

\begin{lstlisting}[caption={Outer loop parallelisation of a nested loop region},label={alg:outerloop}]
#pragma omp parallel for private (j)
for(i = 0; i < I; i++){
  for(j = 0; j < J; j++){
    work();
  }
}
\end{lstlisting}

Parallelising the outermost loop is often a good choice, as it minimises the parallel overheads
of the OpenMP implementation (such as the initialisation of the parallel region, the scheduling 
of loop iterations to threads and the synchronisation which takes place at the end of the
Parallel loops). More extensive work on the overheads of various OpenMP directives
can be found in \cite{Chen:1990:ISG:325164.325150}.

Despite the advantages of the Outermost Loop parallelisation strategy in this context, there 
are drawbacks of this choice. The maximum amount of available parallelism is limited
by the number of iterations of the outerloop loop. Considering the example code in Listing \ref{alg:outerloop}, it
is only possible to have $I$ tasks being executed in parallel. This restricts the number of threads the code 
can utilise upon execution, and therefore the number of processors or cores that can be exploited.

\subsection{Inner loop}
This is a variant on the outermost loop strategy, with the difference that one of the inner loops 
of the region is chosen to be parallelised.  This approach will only be required or beneficial if the
outer loop does not have enough iterations to parallelise efficiently as this variant on the 
parallelisation strategy introduces parallelisation overheads by requiring the parallelisation to be 
performed for each loop of the outerloop rather than once for all the loops (as shown in Listing 
\ref{alg:innerloop}).  Further nesting of the parallelisation (at deeper loop levels) will
further increase the performance problems; the parallel overheads appear a lot more times, whereas the
amount of work of each iteration becomes finer.

\begin{lstlisting}[caption={Inner loop parallelisation of a nested loop region},label={alg:innerloop}]
for(i = 0; i < I; i++){
  #pragma omp parallel for shared (i)
  for(j = 0; j < J; j++){
    work();
  }
}
\end{lstlisting}

Another issue with this strategy is the scenario where loops are not perfectly nested. 
In this situation,
when there are computations in-between the loops, as shown in Listing \ref{alg:poorlynestedloop}, 
parallelising a loop of a deeper level will result in sequential execution of that work. 
Depending on the amount of the execution time which is now serialised, this approach has 
the potential to increase the execution time of the code.

\begin{lstlisting}[float=h,caption={Poorly nested loop region example},label={alg:poorlynestedloop}]
for(i = 0; i < I; i++){
  somework();
  for(j = 0; j < J; j++){
    otherwork();
  }
}
\end{lstlisting}

\subsection{Nested}

The Nested parallelisation strategy exploits the fact that more than one loop can be
executed in parallel. By opening multiple nested parallel regions at different levels
of loops, as presented in Listing \ref{alg:nestedloop}, more threads can be utilised 
during the parallel execution of the code.

Unlike the Outermost Loop and the Inner Loop approaches, which can only utilise
as many threads as the iterations of the loop with the biggest number of iterations,
this strategy can exploit further parallelisation opportunities. Other studies have
shown that nested parallelism can give good results on systems with a large number of
processors\cite{Tanaka00performanceevaluation}\cite{Ayguade:2006:ENO:1143496.1143504}.

\begin{lstlisting}[caption={Nested loop parallelisation of a nested loop region},label={alg:nestedloop}]
#pragma omp parallel for private (j)
for(i = 0; i < I; i++){
  #pragma omp parallel for shared (i)
  for(j = 0; j < J; j++){
    work();
  }
}
\end{lstlisting}

\subsection{Loop Collapsing}

The loop collapsing strategy takes a different approach for exposing additional parallelism
within nested loop regions. By performing code transformations, multiple nested
loops are combined, or collapsed, into a single loop. The newly created loop has a larger 
amount of iterations, which can be distributed to the threads. 

As of version 3.0, OpenMP supports loop collapsing by using the COLLAPSE clause
in the Loop Construct, requiring the programmer to provide the number of loop levels
to collapse.   To be able to use the COLLAPSE clause the loops have to be perfectly 
nested (i.e. no code between the loops) and the number of loop iterations (when multiplied together) 
need to be able to be regularly divided.  Loop collapsing can produce better results than both the inner loop 
and nested loop strategies, since the parallel overheads are minimal, however it is not always 
available, either because not all compilers support OpenMP version 3.0, or because the conditions 
outlined above cannot be met.

\begin{lstlisting}[caption={Parallelisation of a nested loop region with loop collapsing},label={alg:collapsedloop}]
#pragma omp parallel for collapse (3)
for(i = 0; i < I; i++){
  for(j = 0; j < J; j++){
    for(k = 0; k < K; k++){
      work();
    }
  }
}
\end{lstlisting}

\subsection{Loop Selection}
OpenMP already provides a way of forcing a parallel region to execute sequentially
with the use of the {\tt if} clause on OpenMP directives.  The {\tt if} clause, of the following form, 
$if(scalar-expression)$, is used to determine at runtime whether the code enclosed in the 
parallel region should execute sequentially or in parallel. When the scalar expression of 
the clause evaluates to 0, the region is executed sequentially. Any other value will result 
in parallel execution.   However, a new parallel region is always created in either case. 
The presence of the {\tt if} clause only affects the number of threads that get assigned to the parallel region. 
When sequential execution is triggered, the code is only executed by the master thread, for 
parallel execution all threads execute the code.

Furthermore, with the {\tt if} clause, programmers are still required to manually write code which makes the decision, 
construct sensible scalar-expressions to be evaluated, and manually parallelise each loop that 
is a potential target for parallelisation.

\section{Dynamic loop parallelisation}
\label{sec:dynamicloop}
One of the motivators for this work was a parallelisation that was undertaken of a finite-volume cell-centred 
structured Navier-Stokes code for undertaking Computational Fluid Dynamics (CFD) simulation.  It is a
structured mesh, multigrid, code which works with multiblock grids, and includes a range
of CFD solvers including; steady state, time-domain dual time-stepping, frequency-domain
harmonic balance, and time-domain Runge-Kutta. The general pattern for the computations within the code is 
shown in Listing \ref{alg:exampleloops}.  Whilst this type of computational pattern is not uncommon for 
scientific codes one of the challenges in the parallelisation is that as the code can use a range of different methods, 
as previously outlined, the range of these loops can vary. For instance when performing a time domain simulation the {\it harmonic}
loop has a single iteration. However, when performing a harmonic balance simulation
it can have a range of values, generally between 2 and 16. Furthermore, it is not uncommon
to run large simulations with a single block, or a small number of blocks, meaning
that the {\it block} loop has a very small number of iterations. Finally, each block in the simulation can
have different values for its dimensions.  

In theory, the loop collapsing strategy would be ideal 
for this type of simulation code as this would enable parallelisation without having to deal with the 
varying sizes of the nested loops.  However, it cannot be guaranteed that for all input datasets the loop
iterations can be regularly divided, and there are also particular areas of the code where the loops are not 
perfectly nested.

\begin{lstlisting}[caption={Example scientific code loops},label={alg:exampleloops},basicstyle=\scriptsize\ttfamily]
for(iter = 0; iter < n_iters; iter++){
  for(block = 0; block < n_blocks; block++){
    for(harmonics = 0; harmonics < n_harmonics; harmonics++){
      for(j_cell = 0; j_cell < n_cells_j; j_cell++){
        for(i_cell = 0; i_cell < n_cells_i; i_cell++){
          perform computations;
        }
      }
    }
  }
}
\end{lstlisting}

Given the different techniques that can be used to parallelise nested loops, the occurrence 
of nested loops in many scientific simulation codes, and the fact that the loop iterations of 
nested loops can change for different input datasets of a code or when performing different functions 
with a code, we wanted a system that enabled the selection of different parallelisation choices to be 
available to code at runtime when the specific ranges of the nested loops are known.

Our strategy for providing this functionality is to create code, based on the provided user code, that can 
perform a parallelisation of any of the nested loops and add decision making algorithms to dynamically choose, at runtime, 
which parallelisation is used.  Specifically, we have created tools that create multiple versions of a loop
within a region of nested loops in order to make a dynamic choice on whether a loop
should execute sequentially or in parallel

In general, code duplication is considered bad programming practice as it can, amongst other issues, lead to update 
anomalies (where not all instances of the functionality are modified when modifications occur) and thus 
damage the maintainability of the code. However, if the duplicate code (in our instance the serial and parallel
versions of each loop in the nested loop structure) can be generated automatically for standard user 
code then it will not adversely affect the maintainability of the user program.

We created a source-to-source compiler that recognises compiler directives within user's 
source code and uses them to pre-process the source code and generate a program that has the alternative 
parallelisation strategies encapsulated within it.  By exposing a simple interface to the programmers 
through compiler directives, which are similar to the already familiar OpenMP compiler directives, we can 
automatically provide the dynamic parallelisation functionality for users without requiring significant 
changes to the original source code.  Furthermore, this approach provides the users the choice of enabling 
or disabling our functionality with minimum effort.

To complement the code duplication we have also implemented functionality (in a small runtime library) that produces 
the code which is responsible for deciding what parallelisation to perform automatically.  
The decision functionality considers the number of iterations of a loop in order to chose a parallelisation
strategy that makes best use of the processors or cores available.  Our implementation is currently limited 
to parallelising a single loop of a nested loop region, taking advantage only the Outermost and Inner loop strategies.

Other authors \cite{Duran04runtimeadjustment} have already taken a similar approach by modifying the OpenMP 
runtime library in order to make these decisions dynamically. However, applying this logic in the OpenMP
runtime library would have limited the implementation to a specific compiler. Using our source-to-source 
compile approach we are aiming to transfer the logic in user code in order to maintain
the portability of our solution.

In addition to simple heuristics, we also explored the idea of a profile-based approach at
runtime in order to detect the best possible parallelisation strategy with time measurements.
A heuristics based approach alone cannot capture any information on the amount
of the actual computations when making a decision on parallelising a loop. Whilst this is generally 
irrelevant for perfectly nested loops (as all the work is in the lowest loop), it may have more 
of an impact where there is work between the different loops as well.  There may also be situations 
where a different inner loop has slightly more iterations than an outer loop so could be chosen by a 
simple heuristic as the place where the parallelisation occurs but the overheads associated with parallelising 
that inner loop actually make this a suboptimal choice.  Providing a profiling based decision mechanism 
may help with both these scenarios, and enable us to identify situations where, for instance, using less threads 
to parallelise an outer loop might provide a better execution time. The idea of an auto tuning code has already been 
proposed by other compiler-related researches \cite{Hall_looptransformation}
\cite{Pluto_auto} for producing optimised code, we apply similar logic.

\section{Source-to-source compiler}
\label{sec:s2scompiler}
\begin{figure}[!t]
\centering
\includegraphics[width=2.5in]{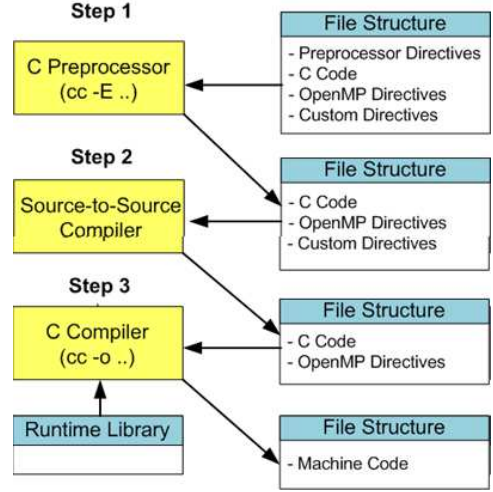}
\caption{Compilation process using the source-to-source compiler}
\label{fig:compileseq}
\end{figure}

Our source-to-source compiler acts as a preprocessor to C code which can contain
OpenMP directives, as well as our own directives.  The compiler parses the code, 
and creates an internal representation of the code in the form of an Abstract Syntax Tree (AST). 
The regions of the input code that contain our directives are translated into the semantics 
of the C programming language and OpenMP directives during the parse phase, 
and appropriate nodes for these regions are placed in the AST. The created AST 
is then translated back to C code with OpenMP directives.  This generated code 
is then compiled using a standard, OpenMP enabled, C compiler to produce a 
parallel executable (this process is illustrated in Figure \ref{fig:compileseq}.

Our compiler, implemented using the Lua\cite{Lua} programming language along with 
the Lpeg\cite{Lpeg} parsing library, recognises a number of our own bespoke compiler 
directives of the form $\#pragma\ preomp$.  A loop that is preceded by a
$\#pragma\ preomp\ for$ directive is considered by our compiler as a suitable candidate for
applying parallelisation. When such a loop is found, our compiler performs the necessary
code transformations so that a decision can be made at runtime whether the loop
should run sequentially or in parallel (and to ensure that both the sequential and parallel 
versions of the loop are available in the executable at runtime). 
In addition to this, a simple analysis of the loop is performed in order to facilitate the 
computation of a loop's iterations during the making of the decision.
An example of such a code is presented in Listing \ref{alg:preompfor}.

\begin{lstlisting}[caption={A nested loop region with preomp},label={alg:preompfor}]
#pragma preomp parallel for private(j)
for(i=0; i<I; i++){
  #pragma preomp parallel for shared(i)
  for(j=0; j<J; j++){
    work();
  }
}
\end{lstlisting}

Furthermore we also extend the grammar to support an additional clause, the $parallel\_
threshold(expression)$ clause. This is optional, and when it is not present the
compiler will assume a default value of 1.0.  This clause is used to allow 
control over when a loop is parallelised, and will be discussed further in Section \ref{sec:decfuns}. 

\subsection{Code Duplication}
The main function of the source-to-source compiler is to take the original user code 
and duplicate the loops to be parallelised so that there are both serial and parallel 
versions of those loops that can be selected at runtime.  As previously mentioned our 
system only allows one loop to be parallelised at any given time (although which loop 
is parallelised can change over the runtime of a program as the parameters of the loop 
change), but both the serial and parallel versions of all the loops to be parallelised 
must appear in the executable to enable a selection at runtime to take place.

When a loop is preceded by a $\#pragma\ preomp\ for$ directive, 
the loop is duplicated and wrapped in a normal $if-else$ statement which evaluates a 
decision function from our runtime library and selects the $if$ or $else$ branch based on the 
outcome of the evaluation.

\subsection{OpenMP if}
As a comparison to our code duplication approach we also implemented the same functionality 
uses the existing if clause of the OpenMP Parallel Construct.  Our custom directive is translated 
into an OpenMP Parallel For directive, with an attached {\tt if} clause in order to decide whether 
to execute the loop in parallel or not (rather than a serial and parallel version of the loop). 
The expression of the {\tt if} clause consists of a call to a decision function of our runtime library, 
which takes the evaluated expressions of the loop's information in order to make a decision.  
This functionality was included to allow a comparison of our approach to the standard method that 
developers could currently use to provide dynamic selection of parallelism with OpenMP.

\begin{figure}[!t]
\centering
\includegraphics[width=2.5in]{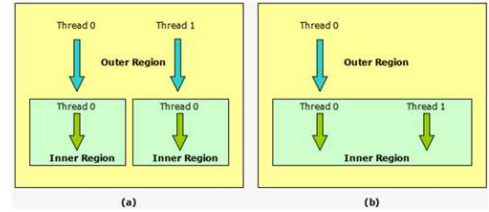}
\caption{An example of using the if clause to parallelise (a) the outer and (b) the inner
loop of two nested loops with two threads}
\label{fig:openmpif}
\end{figure}

However, a major drawback of this approach (and the reason we do not uses it for our 
functionality) is that a parallel region will be created  regardless of whether a loop is parallelised 
or not. Considering the example in Figure \ref{fig:openmpif}, parallelising the outer loop of two nested 
loops with two threads will result in three parallel regions. Each thread of the outer region will 
create a new parallel region and become its master. In the case of the inner loop being parallelised, two parallel
regions are created. For nested regions with a larger number of loops this method has
the potential to produce excessive parallel overheads.

\section{Decision functions and the runtime library}
\label{sec:decfuns}
The runtime library implements the logic for deciding which version of a loop is chosen during
execution. Once a code has been processed by the source-to-source compiler it must then be 
linked with our runtime library to enable this functionality to be used.

\subsection{Decision Based On Heuristics}
Here we use heuristics, based on information collected at runtime, to decide whether a loop 
should execute sequentially or in parallel. The idea of this approach is to look for the first 
loop that has enough iterations to utilise all of the available threads, based on the assumption 
that parallelising outer loops is more efficient than parallelising inner loops as the amount of 
parallel overheads should be lower (as the OpenMP parallel regions are encountered less frequently). 

Before the execution of a loop, the decider checks whether a loop of
an outer level is already running in parallel. If this condition is met, then the loop is
serialised. In the case that no outer loop is running in parallel the number of iterations
of the loop is calculated and it is divided with the available number of threads. If this
results in a value that is greater than or equal to a specified threshold, then the parallel
version of a loop is chosen, otherwise the loop is serialised. As discussed in 
Section \ref{sec:s2scompiler}, the default value of the threshold is 1 (there must be no idle threads) although this 
can be controlled by the user.

The calculations of the iterations is based on the parameters of the loop which are
extracted by the source to source compiler and are provided as arguments to the decision
function. In the case that the original code of the loop uses variables for its boundaries,
any change in their value will also be captured by the decision function during the
calculation. This design allows constant monitoring of any changes in the iterations of
the loops which also results in dynamic adaptation of the parallelisation strategy during
the execution of the program.

The algorithm is very simple and with minimum overheads. Moreover, there is no need
to maintain any state for the loops. However, the logic which is used by the function
is based on optimism. It only considers the amount of parallelism exposed by the loop
regardless of whether the amount of work of the loop is big enough to justify any overheads of 
the parallelisation or whether there is any work between loops.

\subsection{Decision Based On Heuristics With Profiling}
To address the potential issue with the basic decision based on heuristics previously discussed 
we also implemented a more complex decision function based on both the size of loops and some evaluation 
of the work in the loops. In the same manner as the heuristics decider, it uses the same information 
extracted by the source to source compiler in order to determine whether the loop should be 
parallelised or not. However, if a loop does not meet the conditions, then the function reverts 
to a profiling mode in order to decide which version of the loop, serial or parallel, to 
choose from based on timings.

\begin{figure}[!t]
\centering
\includegraphics[width=2.5in]{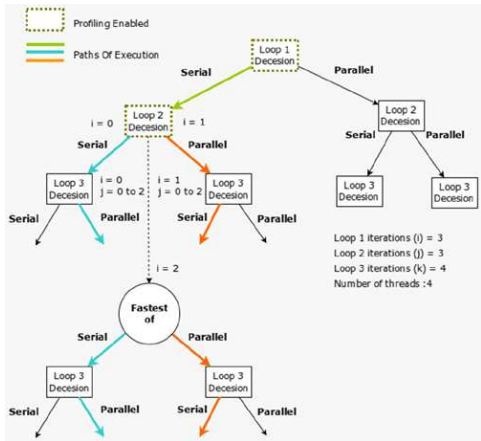}
\caption{An example of the Heuristics With Profiling Decider on three loops}
\label{fig:profilingheur}
\end{figure}

The first time a loop is executed, the heuristics decider determines if the loop should 
be parallelised. If the conditions are not met, the sequential version of the loop is 
chosen and profiling is enabled for this loop. At the next execution of the loop, the 
evaluation of the heuristics is still performed. If the conditions are still not met (for example there where no changes
in the iterations of the loop), the loop is now parallelised since at this point we only have timing
information for the serial version. Consecutive executions of the loop will first check the heuristics
conditions, falling back to profiling mode if the condition is not satisfied.
However, the function will detect that timings for both versions are available and utilise the information 
gathered from profiling to decide what loop to parallelise (providing the number of iterations of the loop 
have not changed), with the fastest version chosen as the final decision. In contrast to
this, if the amount of work is not the same (i.e. the number of loop iterations has changed) 
the timings get invalidated, and profiling is re-initiated.

To implement this functionality requires additional code, when compared to the basic heuristic decision 
function.  This will impose an extra overhead to the produced program, although if the loop iterations 
are static throughout the run of a program the profiling overhead will only be imposed in the first few 
iterations of the program.  Figure \ref{fig:profilingheur} outlines this with an example of three nested loops.

\section{Performance Evaluation}
To evaluate the performance of our new functionality we aimed to benchmark it against standard,
static, OpenMP parallelisations with a range of different configurations.  In particular, we 
focussed on varying the number of loop iterations, the amount of work between and within loops, 
and the number of changes that occur to loop bounds during execution to evaluate whether and when 
our approach is beneficial compared to a static parallelisation.

To undertake these benchmarks we used two different codes.  The first is a synthetic, configurable, benchmark 
C code, shown in Listing \ref{alg:synthbench}, which we constructed for this evaluation.  The number of 
iterations of each loop can be configured, as can the amount of work that is simulated (by calling the $delay$ 
function) between the second and third loops, and within the third loop.

\begin{lstlisting}[caption={Synthetic benchmark code},label={alg:synthbench}]
for(i=0; i<num_iters; i++){
  for(j=0; j<outer_iters; j++){
    delay(outer_delayreps);
    for(k=0; k<inner_iters; k++){
      delay(inner_delayreps);
    }
  }
}
\end{lstlisting}

The second benchmark code was an extract from the CFD code outlined in Listing \ref{alg:exampleloops}.  This 
code is more complex than the synthetic benchmark and more representative of realistic scientific simulation 
codes.  This code is used to explore the performance of our solution when the loop iterations vary and when 
the bounds of loops are dynamic during the course of the execution of the benchmark (i.e. one or more loops 
change their loop bound as the outer loops are progressed).

\subsection{Benchmark Environment}
The platform used to evaluate the dynamic loop parallelisation functionality was Ness\cite{ness}, at EPCC.
The system is composed by two parts, a front-end for development and job submission
and a back-end for job execution. The management of the two parts is handled by
the Sun Grid Engine which allows submission of jobs from the front-end that must be
executed on the back-end nodes in isolation.

The back-end part of the system is composed by two SUN X4600 Shared Memory nodes.
The central processing unit (CPU) of each node is an AMD Opteron processor of 16
2.6GHz processing cores and 32 GB  or main memory. Each core
has 64K of L1 cache for data and 64K L1 cache for instructions. In addition there is also
1 MB of L2 available to each core (combined for data and instructions).

We used the Portland Group (PGI) C compiler for the majority of the benchmarks, with the following 
compiler flags: {\bf-O4},{\bf-c99},{\bf-mp}.  For the benchmarking involving the OpenMP $if$ 
functionality we used the GNC C compiler instead as the version of the PGI compiler we used does not 
support a thread team of a nested parallel region to have more than one threads when an outer region is serialised
with the if clause (this seems contrary to the OpenMP specification where the $if$ clause only
affects the number of threads that get assigned to a particular parallel region, not the
thread teams of its nested regions).  When using the GNU C compiler we used the following compiler flags: 
{\bf-O3},{\bf-stf=c99},{\bf-fopenmp}.

Timing information was collected using the $omp\_get\_wtime()$ function, with each benchmark executed three 
times and the worst time taken (since this is the limiting factor for the execution time).

\subsection{Synthetic benchmark results}

If we consider the example code in Listing \ref{alg:synthbench}, the execution time of the code of the two
internal nested loops when only the outer loop is parallelised with a certain amount of
threads ($outer\_threads$) can be calculated as shown in Equation \ref{equ:tpouter}. $T_{p_{Outer}}$ is the
execution time when parallelising the outer loop, $T_{outer\_work}$ is the time needed for the
work in-between the loops and $T_{inner\_work}$ is the time needed for the amount of work
within the innermost loop.

\begin{IEEEeqnarray}[basicstyle=\tiny\ttfamily]{ll}
T_{p_{Outer}} = \frac{outer\_iters*(T_{outer\_work} + (inner\_iters * T_{inner\_work}))}{outer\_threads}
\label{equ:tpouter}
\end{IEEEeqnarray}

In a similar fashion, when parallelising the inner loop using $inner\_threads$, the execution
time of the loops is shown in Equation \ref{equ:tpinner}.

\begin{IEEEeqnarray}[basicstyle=\tiny\ttfamily]{ll}
T_{p_{Inner}} = outer\_iters * (T_{outer\_work} + \frac{inner\_iters * T_{inner\_work}}{inner\_threads})
\label{equ:tpinner}
\end{IEEEeqnarray}

If we want to have a reduction in the overall execution time by parallelising the inner
loop, the constraint $T_{p_{Inner}} < T_{p_{Outer}}$ must be satisfied. Solving this constraint in terms
of $T_{outer\_work}$ we can get the maximum allowed threshold of the execution time for
the work of the outer loop as shown in Equation \ref{equ:maxwork}. It is worth mentioning that 
this model is an ideal performance model, where the work is evenly distributed to the threads. In reality, the
time of $T_{outer\_work}$ might be affected by the presence of parallel overheads.

\begin{IEEEeqnarray}[basicstyle=\tiny\ttfamily]{ll}
T_{outer\_work} < \nonumber\\ \frac{inner\_iters * T_{inner\_work} *  ( \frac{1}{outer\_threads} - \frac{1}{inner\_threads} )} 
{1 - \frac{1}{outer\_threads}}
\label{equ:maxwork}
\end{IEEEeqnarray}

\begin{figure}[!t]
\centering
\subfloat[Outer Loop Work : 0s]{\includegraphics[width=0.25\textwidth]{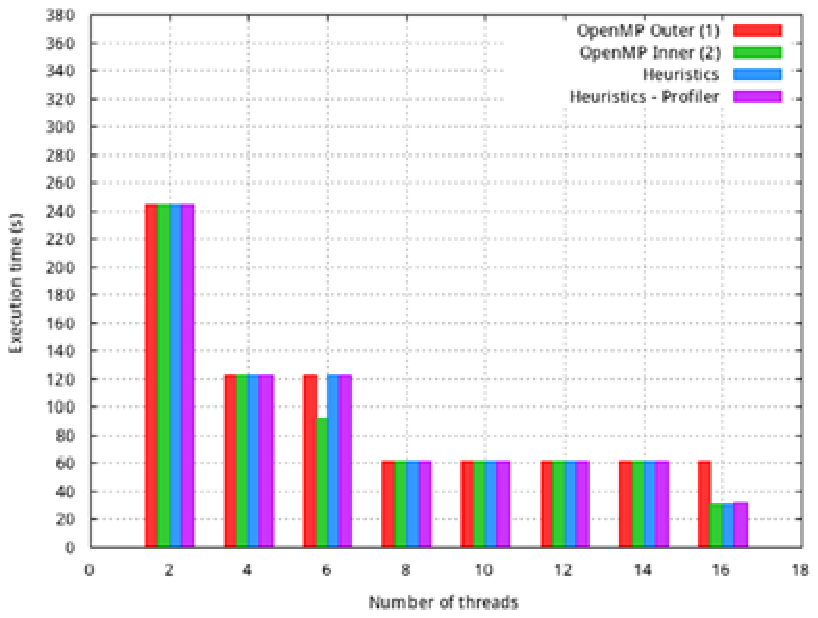}\label{fig:syntheticbench0}}
\subfloat[Outer Loop Work : 0.022s]{\includegraphics[width=0.25\textwidth]{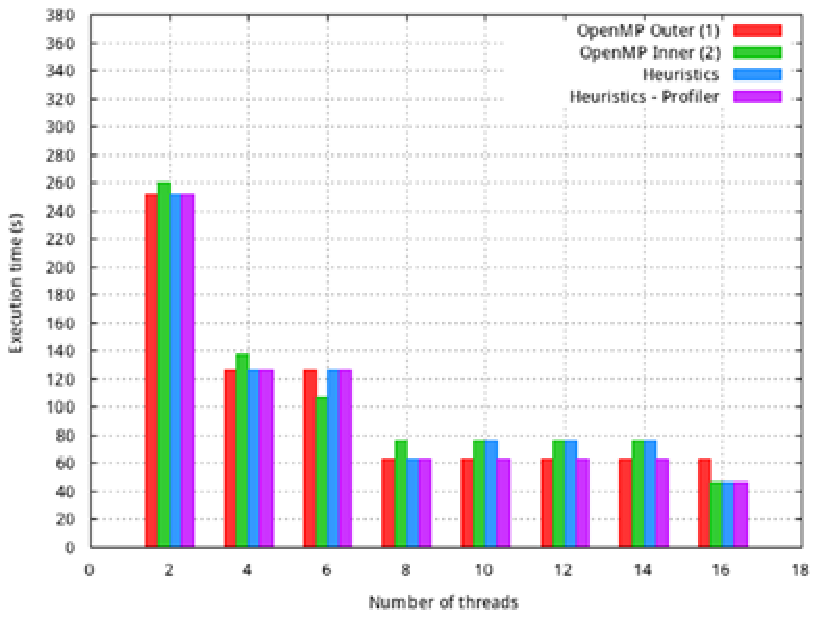}\label{fig:syntheticbench0022}}  \\
\subfloat[Outer Loop Work : 0.079s]{\includegraphics[width=0.25\textwidth]{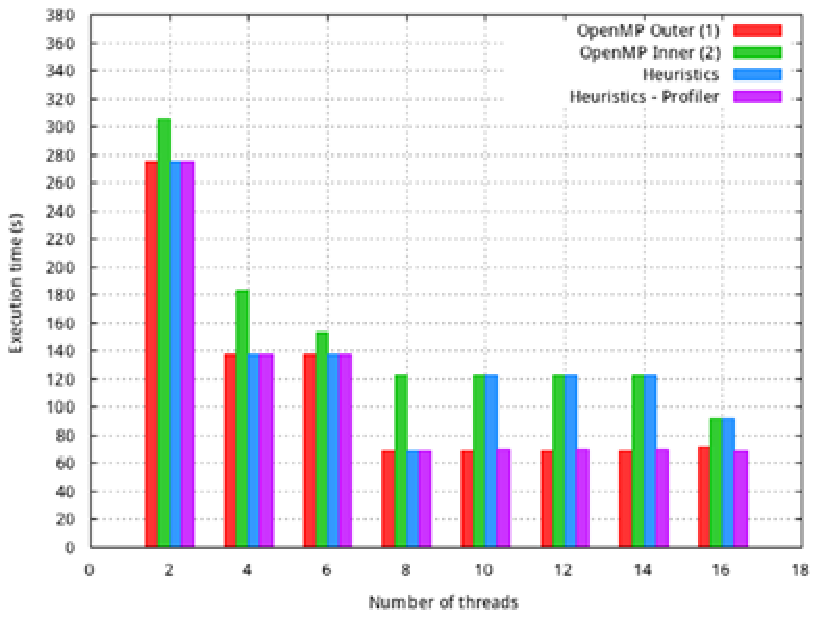}\label{fig:syntheticbench0079}}
\subfloat[Outer Loop Work : 0.15s]{\includegraphics[width=0.25\textwidth]{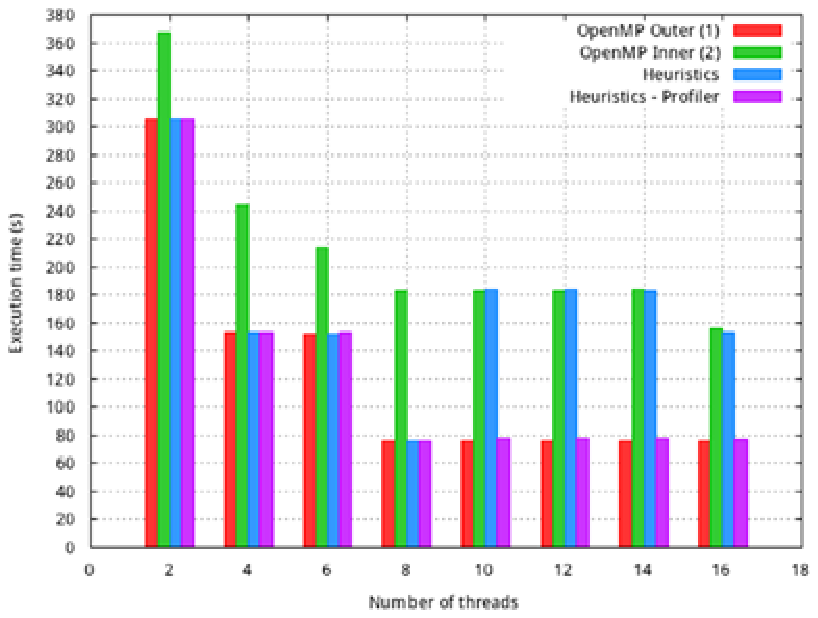}\label{fig:syntheticbench0150}}
\caption{Synthetic Benchmark results with varying levels of work between the loops}
\label{fig:syntheticbench}
\end{figure}

In order to test our hypothesis, we measured the amount of time which is required by
the delay function for various values, with the results shown in Figure \ref{fig:syntheticbench}.  The graphs 
in Figure \ref{fig:syntheticbench} show the performance of four different parallelisation strategies. 
$OpenMP\ Outer(1)$ and $OpenMP\ Inner(2)$ are the results from manual, static, parallelisations of the individual loops in the 
benchmark.  $Heuristics$ are the results from our basic decision function using a value of one (i.e. only 
parallelise the loop if there are more iterations than threads available), and $Heuristic\ \-\ Profiler$ are 
the results from our system using the profiling functionality where appropriate.

From the results it is evident that when the loops are perfectly nested, and regular (i.e. the loop bounds are not changing), then 
there is no benefit from using the profiling functionality.  The basic heuristics will choose the optimal loop to parallelise 
apart from when we are using 6 threads.  The variation in outcomes for 6 threads is is a consequence of the number of loop iterations 
chosen for the benchmark (8 iterations of the outer loop and 16 iterations of the inner loop). The distribution
of 8 iterations to 6 threads results in all of the threads to get assigned 1 iteration of the
outer loop each, and 2 of the threads get and extra iteration. The total
execution time in this case is limited by the slowest threads, which is the time of 32
iterations; 2 iterations of the outer loop multiplied by 16 iterations of the inner one.
Parallelising the inner loop with 6 threads however, 2 of the thread get from 2 iterations
whereas the rest the threads get 3 iterations each. In this case, the total execution
time of the parallel loops is the amount of time required for 24 iterations; 3 iterations of
the inner loop multiplied by 8 iterations of the outer loop. Since both decision functions
only utilise the heuristics decision (when the number of threads is less than the number of 
iterations) they cannot exploit this opportunity as no profiling is actually performed in this case.  This could 
be altered by setting the decision heuristic to a value other than 1 (i.e setting the heuristic to 1.5).

From the graphs we can observe that our threshold value calculations hold.  For the parameters we used 
for this benchmark the calculated threshold value is approximately $T_{outer\_work} < 0.0468$ seconds.  
When the work of the outer work is less than the calculated threshold (Figures \ref{fig:syntheticbench0}
\ref{fig:syntheticbench0022}) parallelising the inner loop with 16 threads is still faster 
than parallelising the outer loop with 8 threads.  As the amount of work increases, the impact on the
execution time when parallelising the inner loop is increased, since more work is
now being serialised. In these cases, the heuristics decider makes the wrong choice 
(Figures \ref{fig:syntheticbench0079} and \ref{fig:syntheticbench0150})
since its decision only concerns the amount of iterations of the loops and the available
threads. In contrast to this, when profiling is used in the decision function, it correctly
detected that the fastest execution time is achieved by not parallelising the inner loop.
In the case, where the amount of work of the outer loop exceeds the calculated threshold, 
parallelising the inner loop, even with 16 threads, increases the total
execution time. The benefit from using 16 threads to parallelise the inner loop is not
enough to justify the work that is serialised.

\subsection{CFD benchmarking results}
The first benchmark that we performed using the extract from the CFD code was to compare the 
OpenMP {\tt if} clause with our basic heuristic functionality.  We used, as a reference, the timings 
of the manually parallelised the $n\_blocks$, $n\_harmonics$ and $n\_cell\_j$ loops and compare the
execution time of the heuristics decision function for the two code generation modes of
our compiler. In order to avoid cases of the iterations not being evenly distributed to the
threads, we only consider cases of 2, 4, 8, 12 and 16 threads.  The parameters used for the loop 
iterations are shown in Table \ref{tab:cfdparams}, with varying amount of work in the inner loop.

\begin{table}
\renewcommand{\arraystretch}{1.3}
\caption{Loop parameters used for the CFD code benchmarking}
\label{tab:cfdparams}
\centering
\begin{tabular}{||p{0.1\textwidth}|p{0.3\textwidth}||}
\hline
{\bf Parameter} & {\bf Value}\\
\hline
$iters$ & 500 \\
$n\_cell\_j$ & 2496 or 8 \\
$n\_cell\_i$ & 8 or 2496 \\
\hline
\end{tabular}
\end{table}

We also consider cases where blocks do not have the same shape by altering the values
of the $n\_cell\_j$ and $n\_cell\_i$ loops. No alterations indicate that all of the blocks have a
grid shape of 2496x8($j\_cell$ x $i\_cell$). An alteration of 2 means that the first and third 
blocks have a grid shape of 8x2496 whereas the second and fourth blocks have a shape of 2496x8.

\begin{figure}[!t]
\centering
\subfloat[Small work, no alterations]{\includegraphics[width=0.25\textwidth]{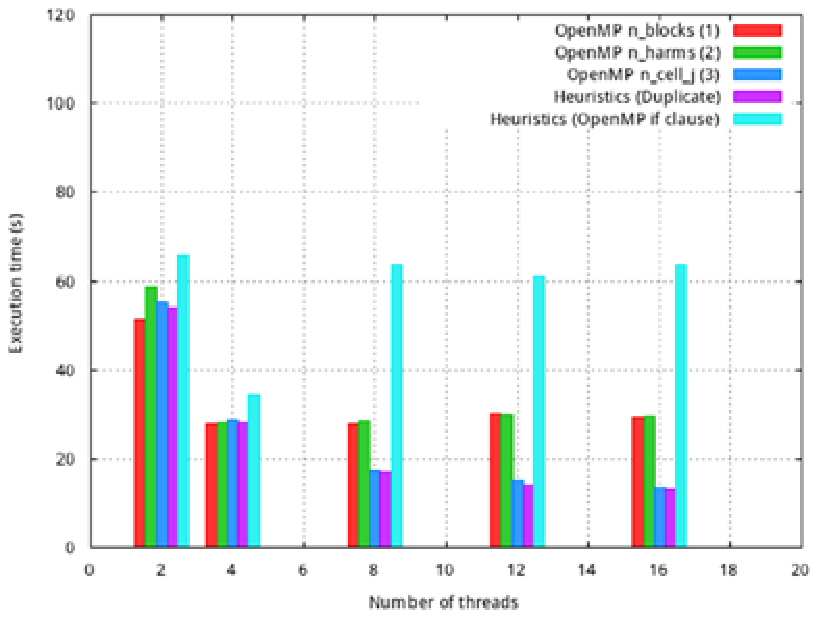}\label{fig:cfdbench440small}}
\subfloat[Small work, 2 alterations]{\includegraphics[width=0.25\textwidth]{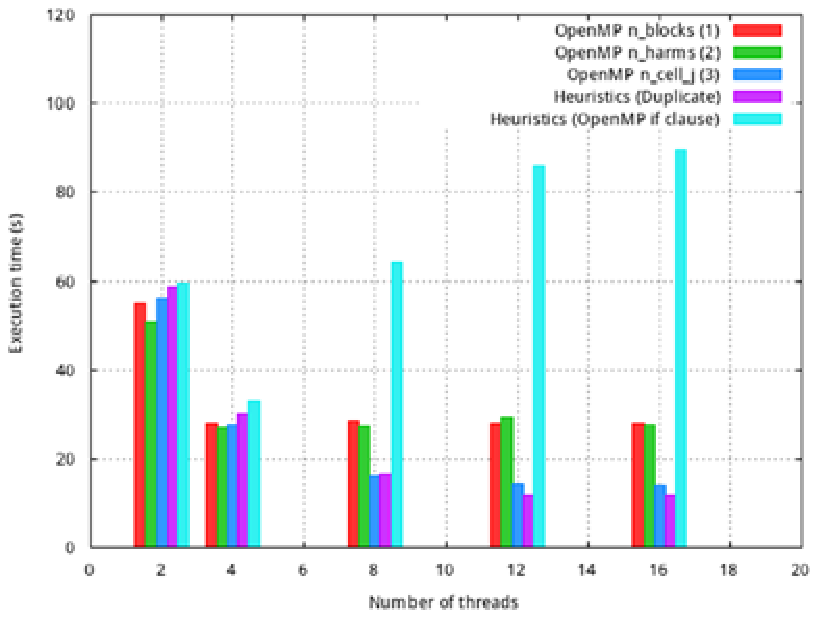}\label{fig:cfdbench442small}}  \\
\subfloat[Large work, no alterations]{\includegraphics[width=0.25\textwidth]{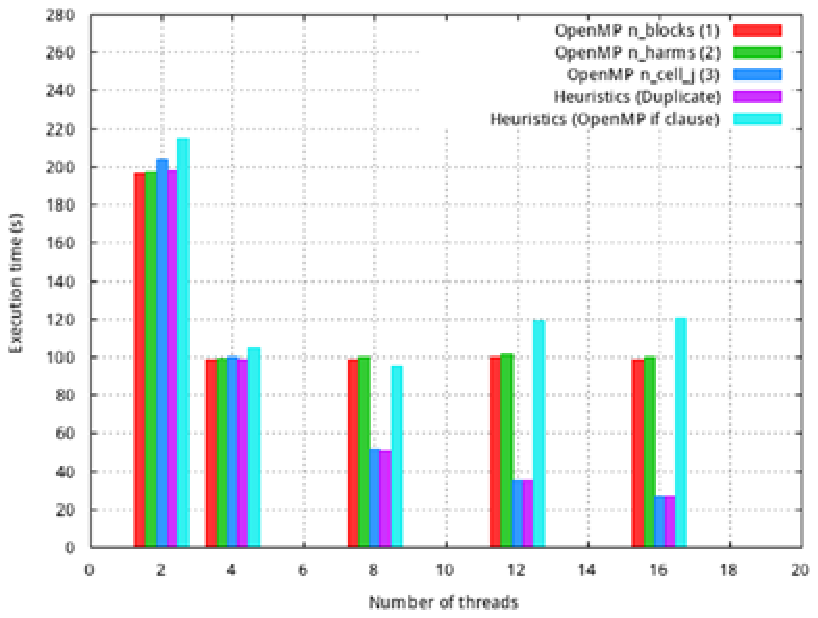}\label{fig:cfdbench440large}}
\subfloat[Large work, 2 alterations]{\includegraphics[width=0.25\textwidth]{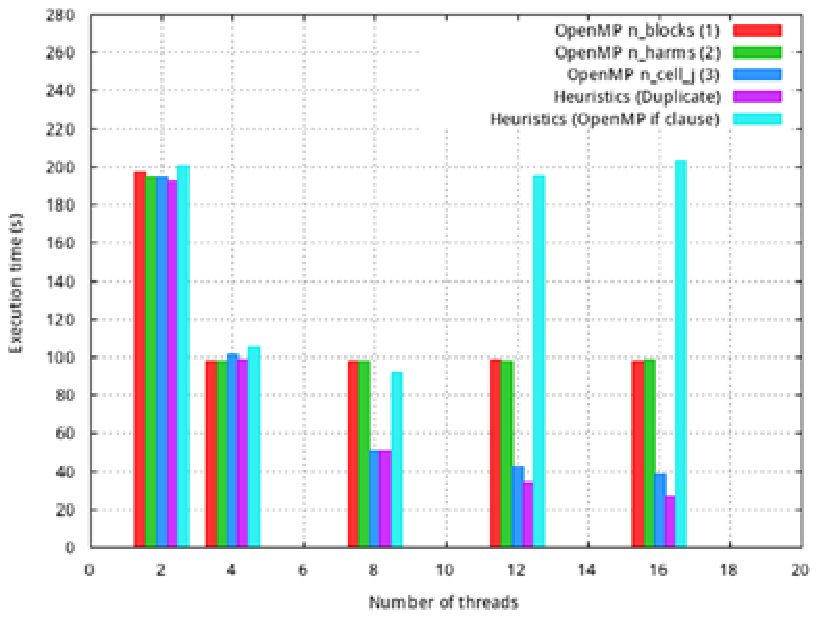}\label{fig:cfdbench442large}}
\caption{CFD benchmark with $n\_blocks$ = 4 and $n\_harmonics$ = 4 with varied alterations in the $i\_$ and $j\_cell$ loops and varied amount of work in the inner loop}
\label{fig:cfdifresults}
\end{figure}

The performance results shown in Figure \ref{fig:cfdifresults} highlight the fact that 
there is a significant difference between our implemented functionality and that provided 
by OpenMP (the {\tt if} clause).  Not only is the {\tt if} clause slower than the basic OpenMP parallelisation, but it also 
increases the overall execution time of the code. For Figure \ref{fig:cfdbench440small}, where 2 and 4 threads are available, only 
the loop of the outer level is parallelised in both code
generation modes. However, the {\tt if} clause mode produces a slower execution time than
the code duplication mode.  When more than 4 threads are used, the parallelisation is applied on the $n\_cell\_j$ loop.
In contrast to the code duplication mode which produces an execution time similar to
the case of statically parallelising the loop, the {\tt if} clause mode is still slower.  A similar performance pattern is seen 
at 16 threads.

Moreover, in the presence of alterations in the shape of the blocks, as shown in Figures
\ref{fig:cfdbench442small} and \ref{fig:cfdbench442large}, the {\tt if} clause mode produces an even slower execution time. 
On the other hand, the code duplication mode can exploit
this opportunity in order to utilise all of the available threads by applying parallelism
on the $n\_cell\_i$ loop.

Increasing the amount of work in the core calculation has a positive effect on
the {\tt if} clause code generation mode. We can observe from Figure \ref{fig:cfdbench440large} that compared 
to Figure \ref{fig:cfdbench440small} the difference between using the {\tt if} clause and the static parallelisation 
is not as large for small numbers of threads.  This is likely to be because the performance cost of executing the {\tt if} clause is proportionally smaller 
compared to the overall execution time. However, the same performance degradation is still observed when 
increasing the number of threads.

The execution times of the code using the OpenMP {\tt if} clause raised some concerns over whether
the code was operating correctly.  After extensive testing and verification we ascertained that 
both versions of the code (the {\tt if} clause and code duplication) were correct and producing the same 
behaviour.  Therefore, we investigated the  parallel
overheads of the OpenMP runtime library of the GCC compiler. 

Other authors \cite{Dimakopoulos:2008:MSO:1789826.1789828} 
have already studied the overheads of nested parallelism on various compilers, including a
more recent version of the GCC compiler than the one used in this work. Their findings
suggest that the implementation of nested parallel regions of the GCC compiler has
significant overheads. What is not presented in their work is whether or not the use of the {\tt if}
clause on nested parallel regions produces the same overheads.  In order to ensure that the behaviour 
we observed in our results is the cause of nested
parallel regions and not the presence of the {\tt if} clause, we have constructed a simple micro benchmark.

\subsection{Nested parallel micro benchmark}

We created four versions of a benchmark code with three nested loops and the delay function of 
the EPCC Micro-benchmark Suite in the block of the innermost loop. The first version of the benchmark creates 
a parallel region on the loop
of the second level. The second version performs the same operation on the innermost
loop. The third version uses the {\tt if} clause on both loops by serialising the outer loop
with a value of 0 and parallelising the inner loop with a value of 1. Finally, the last
version creates a parallel region on both of these loops, however we force the number
of threads on the thread team of the outer loop to 1 using the $num\_threads$ clause. Through this
we manage to reproduce the same behaviour as with the {\tt if} clause code case when
the inner loop is parallelised.

\begin{table}
\renewcommand{\arraystretch}{1.3}
\caption{Micro benchmark results of GNU's C compiler's implementation of nested parallelism}
\label{tab:microresults}
\centering
\begin{tabular}{||p{0.25\textwidth}|p{0.15\textwidth}||}
\hline
{\bf Parallel loop} & {\bf Execution time \newline (seconds)}\\
\hline
Outer & 38.845619 \\
Inner & 153.06809 \\
Nested (with $if$ clause) & 163.05681 \\
Nested (with $num\_threads$ clause) & 162.85479 \\
\hline
\end{tabular}
\end{table}

The number of iterations of the parallel loops are the same as the number of available
threads. Table \ref{tab:microresults} presents the execution times of each case. We can 
see that parallelising the inner loop with nested parallel regions takes 10 seconds 
longer than parallelising the inner loop manually, even for this small and simple 
benchmark. Moreover, the two versions that 
contain nested parallel regions achieve very similar execution times. From this test we can concluded that  
it is likely that the behaviour we observed from the {\tt if} clause code generation mode is 
affected by the overheads of the implementation of the GCC compiler for nested parallel regions.

\subsection{Decision function benchmarking}

\begin{figure}[!t]
\centering
\subfloat[$n\_blocks=4$, $n\_harmonic=8$, no alterations]{\includegraphics[width=0.25\textwidth]{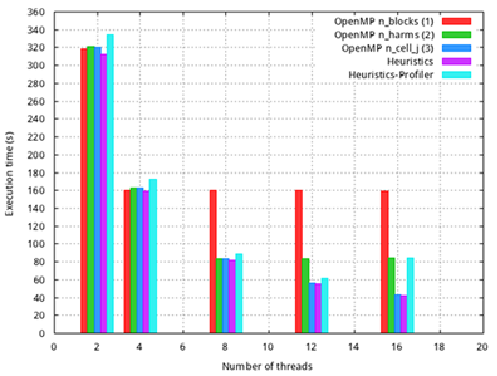}\label{fig:cfddecis480large}}
\subfloat[$n\_blocks=4$, $n\_harmonic=8$, 2 alterations]{\includegraphics[width=0.25\textwidth]{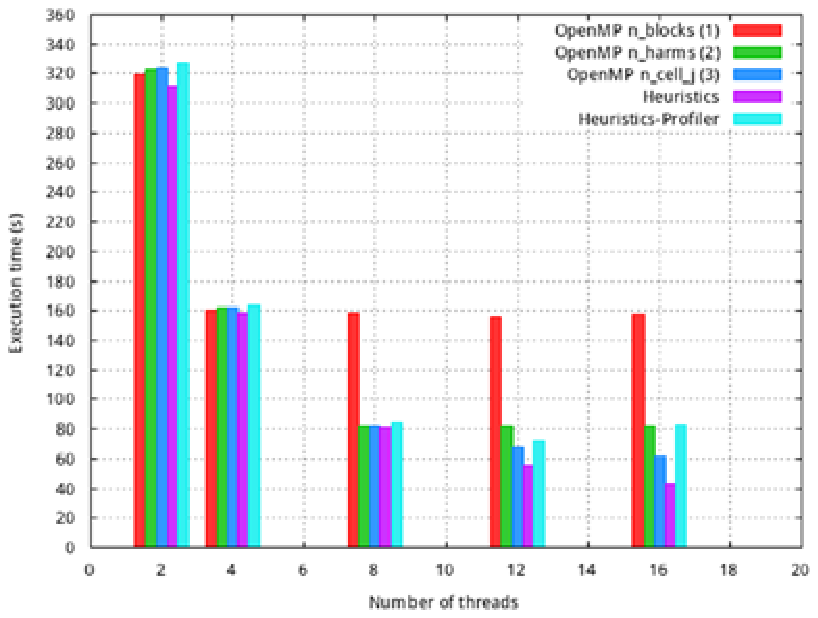}\label{fig:cfddecis482large}}  \\
\subfloat[$n\_blocks=8$, $n\_harmonic=4$, no alterations]{\includegraphics[width=0.25\textwidth]{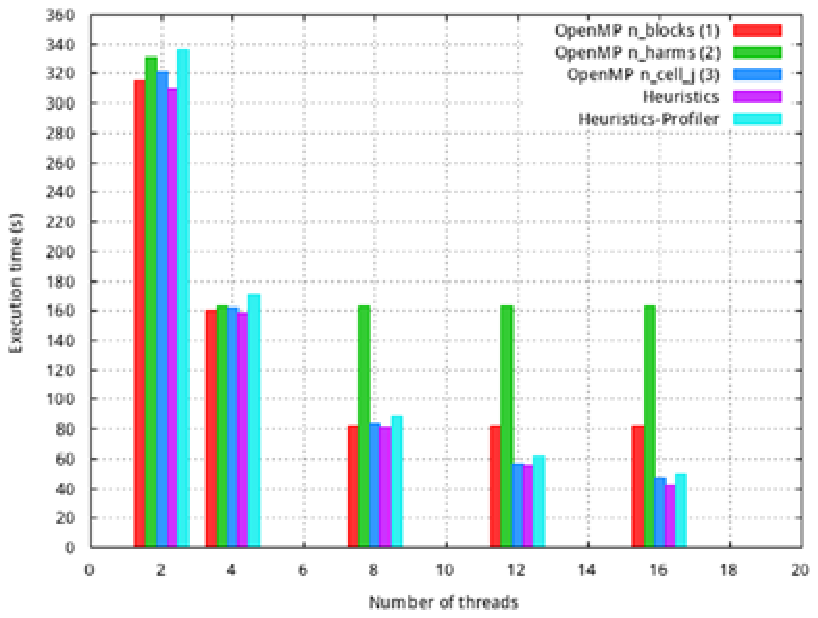}\label{fig:cfddecis840large}}
\subfloat[$n\_blocks=8$, $n\_harmonic=4$, 4 alterations]{\includegraphics[width=0.25\textwidth]{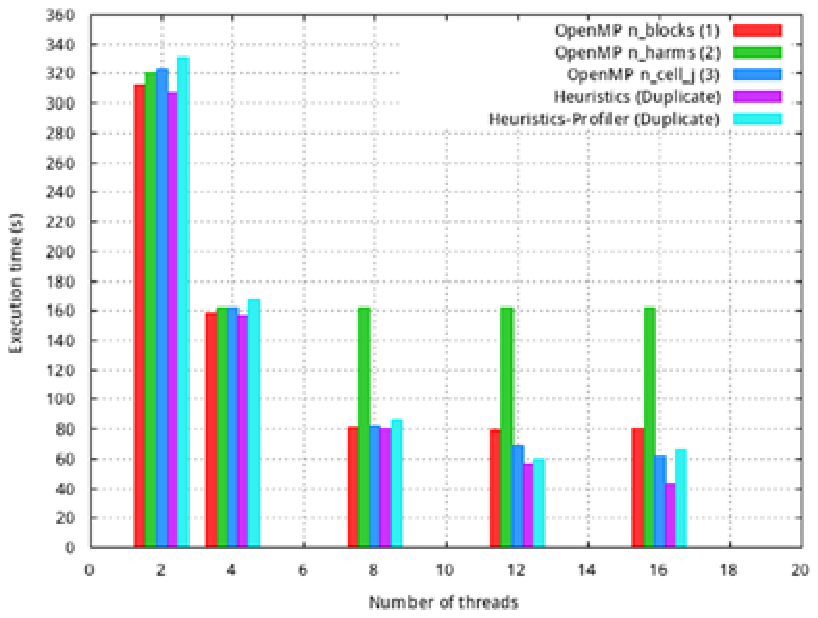}\label{fig:cfddecis844large}}
\caption{CFD benchmark with varied alterations in the $i\_$ and $j\_cell$ loops and a large amount of work in the inner loop}
\label{fig:cfddeciresults}
\end{figure}

Finally, we investigated the performance of our profiling decision functionality for the CFD extract code.
This code is perfectly nested, so the basic heuristic decision function should be optimal here as 
it should chose the best loop to parallelise with very little overheads, whereas the profiling 
function has extra functionality and therefore imposes extra overheads on the performance of the code.
The results from our experiments are shown in Figure \ref{fig:cfddeciresults}.

We can observe from Figures \ref{fig:cfddecis480large} that both the decision functions make the correct choice of parallelisation strategy 
up to 12 threads.  However, the overheads of the profiling functionality have a negative impact on the overall execution time. 
Even when profiling is not actually being performed, the functions which are inserted before and after the execution 
of each loop to count the amount of work performed at each loop level increase the overall time. Moreover,
we can observe that at 16 threads the profiler actually chooses to parallelise the harmonics loop, whereas the heuristics
decider produces the correct behaviour of profiling the $n\_cell$ loop. The timings which are performed for each loop
version during the profiling mode are sensitive to the presence of any overheads which
ultimately affect the decision of the function (such as the overhead of taking the timings).

When alterations
are present in the shape of the loops, as shown in Figures \ref{fig:cfddecis482large} and \ref{fig:cfddecis844large}, the
heuristics decider manages to adapt its behaviour, parallelising the innermost loop
in order to utilise more threads, and can significantly out perform the static parallelisation.

In all of the test cases, the decision function which is based on profiling provides slower
execution times than the decision function which is based on heuristics. Moreover,
the additional logic which is included in the decision function with profiling caused a
suboptimal decision to be made in some situations.

\section{Improved profiling decisions}
The results from the previous benchmarks lead to considerations of the reasons behind
the poor execution of the decision function which performs profiling. Comparing the
functionality of this function with the simple case of the heuristics decision function
there are two sources of additional overheads.

The first one is the logic of profiling each version of a loop. In order to make a choice
between the two versions of a loop, the slow version must also be executed.  However, if 
an actual simulation code runs for a significant amount of time this overhead should 
be negligible (providing the loop bounds do not alter and trigger the profiling functionality 
too many times) as it should only be incurred infrequently.

The second source of overheads is the inclusion of additional function calls before and after
each loop in order to measure the time of the execution and count the amount of work
performed.

The elimination of the functionality for taking the slow path is not possible since this is
the essence of profiling. Both versions of a loop must be executed in order to make a
comparison between their execution time. However, we can relax the conditions on the
validity of the timings.

If we only consider the number of the iterations of the specific loop which is being
profiled, then we can eliminate all the logic that performs the counting of the work for
the internal loops. When the decision function decides that a version of a loop should
be profiled (after the failure of the heuristics conditions) the number of the iterations of
the version of the loop that is going to be executed is saved in the state of the loop at
that point. This way, the code of the function calls which are placed before and after
each loop remains simple, only adjusting the loop level counter of each thread as well as
marking the starting and ending times of the execution of a loop which is being profiled, rather 
than counting the iterations of internal loops as the initial profiling functionality does.

In order to test our theory we have created a new version of the runtime library which
includes the above modifications, called the relaxed profiler.

\begin{figure}[!t]
\centering
\subfloat[Small work, $n\_blocks=4$, $n\_harmonic=4$, 2 alterations]{\includegraphics[width=0.25\textwidth]{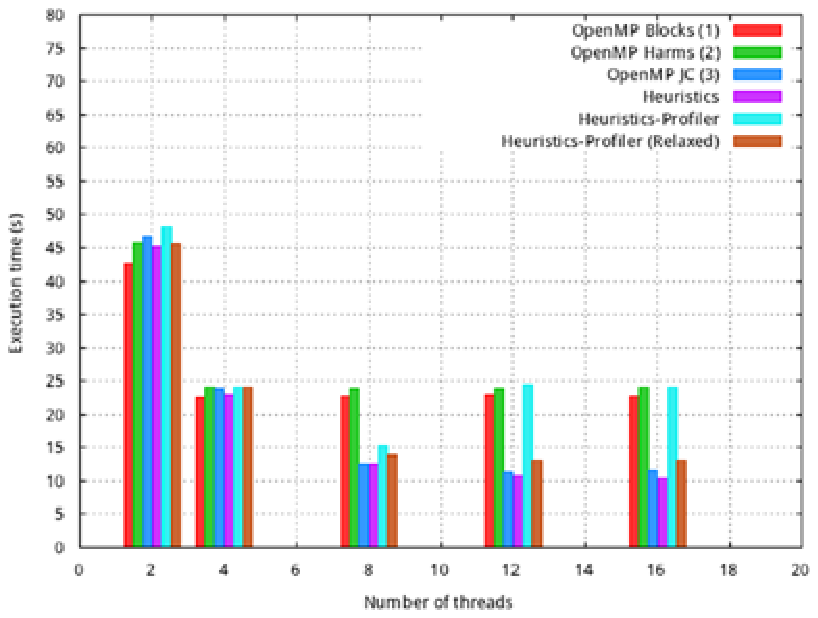}\label{fig:relax442small}}
\subfloat[Large work, $n\_blocks=8$, $n\_harmonic=4$, 4 alterations]{\includegraphics[width=0.25\textwidth]{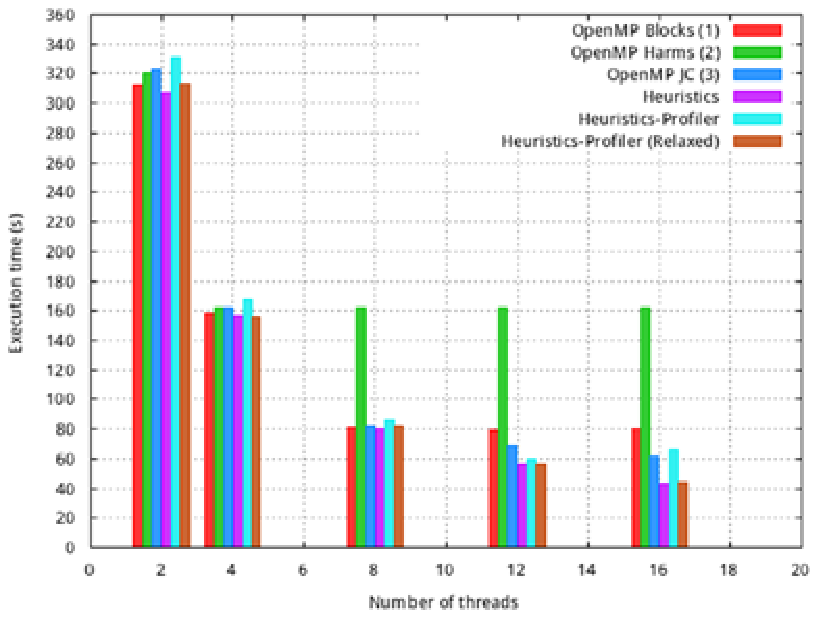}\label{fig:relax844large}}  \\
\caption{CFD benchmark with varied alterations in the $i\_$ and $j\_cell$ loops and a large amount of work in the inner loop}
\label{fig:relaxedresults}
\end{figure}

From the graphs in Figure \ref{fig:relaxedresults} we can see that the removal of the additional 
logic which performs the counting benefits the decision function with profiling.
When no profiling is performed (2 and 4 threads), the relaxed version of the decision
function is faster than the accurate version, and the same performance pattern holds 
when the profiling is performed (8 threads and more for Figure \ref{fig:relax442small} and 
12 threads and more for Figure \ref{fig:relax844large}).

Comparing the execution time of the new version of the decision function with profiling
to the execution time of the heuristics decision function, the latter still produces a faster
execution time, however the difference is not large.
This behaviour is expected, since the presence of profiling introduces additional computations
within the code itself from the functions which are placed before and after
each loop. Moreover, in the cases where the parallelisation is applied on a nested loop,
the decision function must execute both versions of a loop, one of them being the slow
version, in order to make a decision.

Finally, we can see that the relaxed decision function rectifies the problem with the 
original profiling decision function of it choosing the wrong option in some cases.  For 
Figure \ref{fig:relax442small} we can see that at 12 and 16 threads the relaxed profiler 
makes the correct choice, and the same for Figure \ref{fig:relax844large} at 16 threads 
(where the performance of the relaxed profile decision function is comparable to the 
heuristics decision function).

\section{Conclusion}
The main focus of this work was to investigate the possibility of dynamically choosing
at runtime the best loop of a nested loop region which best utilises the available threads.
We have successfully created a source-to-source compiler and a runtime library in order
to automatically allow a dynamic choice to be made at runtime.  As our solution uses a directives
based approach, similar to OpenMP, we requires minimum effort and code change from the 
user's point of view.

We have discovered that the current mechanism users can exploit to perform this, the OpenMP 
{\tt if} clause, does not perform efficiently (at least for the implementation we tested). 
Despite the fact that this behaviour is the
result of the inefficient implementation of the GCC compiler which was used in this
work, the same compiler with the code duplication mode was able to provide additional
speedup in the execution time of the code. From this we conclude that by relying on
the OpenMP runtime library to perform loop nesting, the execution time is limited by
the compiler's implementation of nested parallel regions. Although code duplication
is considered to be a bad programming practice, when it is done automatically, it can
eliminate unnecessary parallel overheads.

We have also shown that some level of auto-tuning (using profiling to select which 
loop to parallelise) can provide performance benefits in certain circumstances, for instance when 
loops are not perfectly nested.

OpenMP is currently generally  used for small scale parallelisation of code, primarily because 
there are very few large scale shared-memory HPC resources.  However, the current trend in 
multi-core processors suggests that in the near future large scale shared-memory resources 
(of order 100-1000s of cores) are likely to be commonly available.  Therefore, shared-memory 
parallelisations are likely to become more utilised and interesting for large scale 
scientific simulations.

\IEEEtriggeratref{9}

\bibliographystyle{IEEEtran}
\bibliography{IEEEabrv,dynamicloop}

\end{document}